\documentclass[11pt]{article}
\usepackage{amsmath,amssymb,color}
\usepackage{amsfonts}
\newcommand{\hoch}[1]{$\, ^{#1}$}

\newcommand{\auth}{H. L\"u\hoch{\dagger\ddagger} and Yi Pang\hoch{\star} }

\def\ft#1#2{{\textstyle{\frac{\scriptstyle #1}{\scriptstyle #2} } }}
\def\fft#1#2{{\frac{#1}{#2}}}
\def\CP{{{\mathbb C}{\mathbb P}}}
\def\0{{\sst{(0)}}}
\def\1{{\sst{(1)}}}
\def\2{{\sst{(2)}}}
\def\3{{\sst{(3)}}}
\def\4{{\sst{(4)}}}
\def\5{{\sst{(5)}}}
\def\6{{\sst{(6)}}}
\def\7{{\sst{(7)}}}
\def\8{{\sst{(8)}}}
\def\sst#1{{\scriptscriptstyle #1}}

\begin{document}

\begin{flushright}
\hfill{ \ }
%MIFP-09-00\ \ \ \ \ \ \ \  }\\
 %\hfill{
%\bf hep-th/yymmnnn}
\end{flushright}

\vspace{25pt}
\begin{center}
{\large {\bf Seven-Dimensional Gravity with Topological Terms}}

\vspace{15pt}

\auth

\vspace{10pt}

\hoch{\dagger}{\it China Economics and Management Academy\\
Central University of Finance and Economics, Beijing 100081}

\vspace{10pt}

\hoch{\ddagger}{\it Institute for Advanced Study, Shenzhen
University, Nanhai Ave 3688, Shenzhen 518060}

\vspace{10pt}

\hoch{\star}{\it Key Laboratory of Frontiers in Theoretical
Physics\\
Institute of Theoretical Physics, Chinese Academy of Sciences,
Beijing 100190}

\vspace{40pt}

\underline{ABSTRACT}
\end{center}

We construct new seven-dimensional gravity by adding two
topological terms to the Einstein-Hilbert action.  For certain
choice of the coupling constants, these terms may be related to
the $R^4$ correction to the 3-form field equation of
eleven-dimensional supergravity. We derive the full set of the
equations of motion.  We find that the static spherically-symmetric
black holes are unmodified by the topological terms.  We obtain
squashed AdS$_7$, and also squashed seven spheres and $Q^{111}$ spaces
in Euclidean signature.

\vspace{15pt}

\thispagestyle{empty}

%\pagebreak
%\voffset=0pt
%\setcounter{page}{1}

%\tableofcontents

%\addtocontents{toc}{\protect\setcounter{tocdepth}{2}}

%%%%%%%%%%%%%%%%%%%%%%%%%%%%%%%%%%%%%%%%

\newpage
%%%%%%%%%%%%%%%%%%%%%%%%%%%%%%%%%%%%%%%%
\section{Introduction}

There has been considerable interest in topological gauge theories
\cite{Deser:1981wh} because of their wide application in physics.
The most studied example is the three-dimensional one.  In addition
to the Einstein-Hilbert term, the theory has the Chern-Simons term,
given by
%%%%
\begin{equation}\label{} S=\fft{1}{\mu}\int d^3x {\rm
Tr}\,(d\omega\wedge\omega+ \ft23\omega\wedge\omega\wedge\omega),
\end{equation}
%%%%%
where $\omega$ can be either a Yang-Mills gauge potential or the
connection for gravity.  Topological Yang-Mills theory can provide a
fundamental interpretation for anyons \cite{Wilczek:1983cy}; it can
also generate Lorentz violation dynamically \cite{Carroll:1989vb}.
Topologically massive gravity \cite{Deser:1983tn} becomes dynamical
with a propagating massive particle, with the mass proportional to the
coupling constant $\mu$. Recently, a cosmological constant is added
and the corresponding boundary conformal field theory (CFT) is
discussed \cite{Witten:2007kt}. The three-dimensional massive
topological gravity is conjectured to be unitary for certain
parameter region even though the theory has higher derivatives in
time \cite{stromingeretal}.

    The attention on higher dimensional generalizations
is considerably less.  The five dimensional Yang-Mills Chern-Simons
term was discussed in \cite{Gunaydin:1984nt}, but there is no
gravity counterpart due to the fact that the holonomy group
$SO(1,4)$ has no invariant rank-3 symmetric tensor. In seven
dimensions, Yang-Mills Chern-Simons terms arise naturally from
${\cal N}=4$ supergravity \cite{Pernici:1984xx}.  As in the case of
three dimensions, we find that such terms in the gravity sector can
be obtained directly from those in the Yang-Mills sector by
replacing the gauge potential $A$ to the connection $\Gamma$.
As we shall see later, these topological terms in seven dimensions
may be related to the anomaly cancelation terms in eleven-dimensional
supergravity.

     In section 2, we present the two topological terms in seven
dimensions, and discuss their properties.  Since they are not
manifestly invariant under general coordinate transformation, we
find it is more convenient to lift the system to eight dimensions in
order to derive the equations of motion (EOMs). We obtain the full
set. In section 3, we construct large classes of solutions.
We find that the static spherically-symmetric black holes are unmodified
by the topological terms.  This is analogous to three dimensions, where
the BTZ black hole remains to be a solution in topologically
massive gravity. In Euclidean signature, we obtain squashed $S^7$ and
$Q^{111}$ spaces.  In particular, one of the squashed seven sphere can be
Wick rotated to become squashed AdS$_7$.  We conclude in section 4.

\section{The theory}

In seven dimensions, there are two topological terms; they are given
by
\begin{eqnarray}\label{A1} S_1
&=&\tilde \mu\int\Omega^{(7)}_1=\tilde \mu\int{\rm
Tr}(\Gamma\wedge\Theta-\ft13\Gamma^3)\wedge{\rm Tr}(\Theta^2)
=\tilde \mu\int\Omega^{(3)}\wedge d\Omega^{(3)},\\
S_2&=&\tilde \nu\int\Omega^{(7)}_2=\tilde \nu\int{\rm
 Tr}(\Theta^3\wedge\Gamma-
 \ft25\Theta^2\wedge\Gamma^3-
 \ft15\Theta\wedge\Gamma^2\wedge\Theta\wedge\Gamma+
 \ft15\Theta\wedge\Gamma^5-\ft{1}{35}\Gamma^7),\nonumber
\end{eqnarray}
with $\Omega^{(3)}={\rm
Tr}(d\Gamma\wedge\Gamma+\ft23\Gamma^3)$. Here, $\Theta$ is the
curvature 2-form, defined as $\Theta\equiv
d\Gamma+\Gamma\wedge\Gamma$, and $\tilde \mu, \tilde \nu$ are
two parameters of
length dimension 5. (We rescale the total action by the
seven-dimensional Newton constant.) The 3-form $\Omega^{(3)}$ has
the same structure as the Chern-Simons term in $D=3$, except that
now $\Gamma$ depends on seven coordinates. $\Omega_1^{(7)}$ and
$\Omega_2^{(7)}$ are topological in the same sense as $\Omega^{(3)}$
being topological in $D=3$. We can lift the system to $D=8$, with
the seven-dimensional spacetime as the boundary.  Then, we have
\begin{equation}\label{domega712}
d\Omega^{(7)}_1=Y_1^{(8)}\equiv{\rm Tr
}(\Theta\wedge\Theta)\wedge{\rm Tr
 }(\Theta\wedge\Theta)\,,\qquad d\Omega^{(7)}_2
 =Y_2^{(8)}\equiv {\rm Tr} (\Theta\wedge\Theta\wedge
 \Theta\wedge\Theta)\,.
\end{equation}
As we have mentioned earlier, these terms can be derived from the
Yang-Mills Chern-Simons terms in \cite{Pernici:1984xx} by changing
the gauge potential to the connection.\footnote{In
\cite{Pernici:1984xx}, the field strength 2-form is
defined by $F=dB+gB\wedge B$, with gauge coupling $g=2$. Then by
rescaling the field $B\rightarrow B/g$ and $F\rightarrow F/g$ and
setting $g=2$, one can obtain the same expressions as
the ones given here.}  Note that the Pontryagin term is proportional
to $Y_1^{(8)} - 2 Y_2^{(8)}$, corresponding to $\tilde\nu=-2\tilde \mu$.
In eleven-dimensional supergravity, there is an $R^4$ correction to the
field equation, namely $d{*F^{(4)}} = \ft12 F^{(4)}\wedge F^{(4)} +
X^{(8)}$, where $X^{(8)}$ is given by %%%%
\begin{equation}
X^{(8)} \propto Y_1^{(8)} - 4 Y_2^{(8)}\,.
\end{equation}
%%%
Thus for $\tilde\nu=-4\tilde \mu$, the topological terms can be obtained
from the $S^4$ reduction of supergravity in $D=11$, and the
coupling constant is proportional to the 4-form M5-brane fluxes.
For large fluxes, this topological term dominates the higher-order
corrections.

    To derive the contribution to the EOMs from the Chern-Simons
terms, it is necessary to perform their variation with respect to
the metric. These topological terms are not manifestly invariant
under the general coordinate transformation, but $Y_1^{(8)}$ and
$Y_2^{(8)}$ are.  We find that a convenient way to derive the
variation is to lift the system to eight dimensions. Let us first
consider the variation of $S_1$. In terms of coordinate components,
we have
\begin{equation}\label{Eq1}
    \int d\Omega^{(7)}_1=\ft1{16}\int d^8x
    \epsilon^{\nu_1\nu_2\nu_3\nu_4\nu_5\nu_6\nu_7\nu_8}
    R^{\mu_1}_{~\mu_2\nu_1\nu_2}R_{~\mu_1\nu_3\nu_4}^{\mu_2}
    R^{\mu_3}_{~\mu_4\nu_5\nu_6}R_{~\mu_3\nu_7\nu_8}^{\mu_4}\,.
\end{equation}
Here we use Greek letters to denote the eight-dimensional
coordinates and Latin letters to represent the seven-dimensional
ones hereafter. We adopt the convention $\epsilon^{12345678}=1$.

   For an infinitesimal variation of the metric $\delta g$, using the
Bianchi identity and the following relation
\begin{equation}\label{}
\delta R^{\mu}_{~\nu\alpha\beta}
=\delta\Gamma^{\mu}_{\nu\beta;\alpha} -\delta
\Gamma^{\mu}_{\nu\alpha;\beta},
\end{equation}
we find that
\begin{eqnarray}\label{Eq2}
\int d\delta\Omega^{(7)}_1&=&-\ft12\int d^8x\sqrt{g}
\Big(\frac{1}{\sqrt{g}}
\epsilon^{\nu_1\nu_2\nu_3\nu_4\nu_5\nu_6\nu_7\nu_8}
    R^{\mu_1}_{~\mu_2\nu_1\nu_2}R_{~\mu_1\nu_3\nu_4}^{\mu_2}
    R^{\mu_3}_{~\mu_4\nu_5\nu_6}
    \delta\Gamma^{\mu_4}_{~\mu_3\nu_7}\Big)_{;\nu_8}\cr
    &\equiv& \ft12 \int d{*J}\,,
\end{eqnarray}
where ``;'' denotes a covariant derivative and $*$ is the Hodge
dual. For simplicity, we have introduced a 1-form current
$J=J_\alpha dx^\alpha$. Its components are given by
\begin{equation}\label{}
    J^{\alpha}=\frac{1}{\sqrt{g}}
    \epsilon^{\nu_1\nu_2\nu_3\nu_4\nu_5\nu_6\nu_7\alpha}
    R^{\mu_1}_{~\mu_2\nu_1\nu_2}
    R_{~\mu_1\nu_3\nu_4}^{\mu_2}R^{\mu_3}_{~\mu_4\nu_5\nu_6}
    \delta\Gamma^{\mu_4}_{~\mu_3\nu_7}.
\end{equation}
Clearly, we have $d{*J}=-\sqrt{g}J^{\alpha}{}_{;\alpha}d^8x$, Thus
we obtain
\begin{equation}\label{}
    \delta\Omega^{(7)}_1=\ft12 {* J},
\end{equation}
up to a total derivative term.  Now restricting the coordinate
indices to seven dimensions only, we have
\begin{equation}\label{}
    \delta S_1=4\tilde \mu\int {\rm Tr }(\Theta\wedge\Theta)\wedge{\rm Tr
    }(\Theta\wedge\delta\Gamma).
\end{equation}
The variation of $S_2$ can be obtained in the same manner, given by
\begin{eqnarray}
  \delta S_2=4\tilde \nu\int{\rm
  Tr}(\Theta\wedge\Theta\wedge\Theta\wedge\delta\Gamma).
\end{eqnarray}
Finally, we make use of the variation of the connection
\begin{equation}\label{}
   \delta\Gamma^{i}_{mj}=\ft12 g^{in}(\delta g_{nm;j}
+\delta g_{nj;m}-\delta
   g_{ml;n}),
\end{equation}
and after integrating by parts, we obtain the contributions to EOMs
from the Chern-Simons terms, given by
\begin{eqnarray}\label{}
    C_1^{ij}&=&\frac{\delta S_1}{\sqrt{g}\delta
    g_{ij}}=\frac{\mu}{4\sqrt{g}}[\epsilon^{ij_1j_2j_3j_4j_5j_6}
    (R^{i_1}_{~i_2j_1j_2}R_{~i_1j_3j_4}^{i_2} R^{jk}_{~~j_5j_6})_{;k}
    +i\leftrightarrow j],\cr
    C_2^{ij}&=&\frac{\delta S_2}{\sqrt{g}\delta
    g_{ij}}=\frac{\nu}{4\sqrt{g}}[\epsilon^{ij_1j_2j_3j_4j_5j_6}
    (R^{k}_{~i_1j_1j_2}R^{i_1}_{~i_2j_3j_4}R^{ji_2}_{~~~j_5j_6})_{;k}
    +i\leftrightarrow j].
\end{eqnarray}

For the total action $S$, which is the sum of the Einstein-Hilbert
action, cosmological constant $\Lambda$ and $S_1 + S_2$, the
corresponding full set of EOMs is given by
\begin{equation}\label{}
    R^{ij}-\ft12g^{ij}R+\Lambda g^{ij}+C_1^{ij}+C_2^{ij}=0.
\end{equation}

    It should be remarked that under a large gauge transformation
$\Gamma\rightarrow\mathcal {O}\Gamma\mathcal {O}^{-1}-
d\mathcal {O}\mathcal {O}^{-1}$, the action transforms as
$S\rightarrow S+\tilde \mu v(\mathcal {O})+\tilde \nu w(\mathcal {O})$,
where
\begin{equation}
v(\mathcal {O}) =\int \ft13 d\Big({\rm Tr} (d\mathcal {O}\mathcal
{O}^{-1})^3\wedge\Omega^{(3)}\Big); \qquad w(\mathcal
{O})=\ft{1}{35}\int {\rm Tr}(d\mathcal
   {O}\mathcal {O}^{-1})^7.
\end{equation}
The $v$ term is trivial and gives no restriction to the parameter
$\tilde\mu$, while the $w$ term should be classified by the seventh
homotopy group of $SO(1,6)$
\begin{equation}\label{}
   \pi_7[SO(1,6)]\simeq\pi_7[SO(6)]\simeq\mathbb{Z}.
\end{equation}
The invariance of $e^{{\rm i}S}$ requires that
\begin{equation}\label{}
   64\pi^4\tilde\nu=2\pi n,~~~~n=0, \pm1, \pm2\ldots.
\end{equation}
This result is completely different from that in three dimensions,
where the $SO(1,2)$ is homotoplically trivial and the mass parameter
is not quantized. Moreover, since $\tilde\nu$ is quantized, $S_2$ will
not be renormalized in the quantum theory.  This suggests some
intriguing properties in the corresponding CFT dual.

\section{Solutions}

\noindent\underline{\bf Spherically-symmetric solutions:}
\medskip

Having obtained the full set of EOMs for topological gravity in
seven dimensions, we are in the position to construct solutions. It
is clear that the maximally-symmetric space(time) is unmodified by
the inclusion of the topological terms.  The next simplest case is
to
consider the spherically-symmetric ansatz, given by %%%%
\begin{equation}
ds^2 = -F(r) dt^2 + \frac{dr^2}{G(r)} + r^2 d\Omega_5^2\,.
\end{equation} %%%%
We find that for this ansatz, the contributions from the topological
terms $C_1^{ij}$ and $C_2^{ij}$ vanish identically.  This implies
that the previously-known static (AdS) black holes, charged or
neutral, are still solutions when the topological terms are added to
the action. This is analogous to three dimensions, where the BTZ
black hole is still a solution in massive topological gravity.
However the thermodynamic quantities such as the mass and entropy
will acquire modifications \cite{Deser:2003vh,Tachikawa:2006sz}.

      As we shall discuss presently, there also exist squashed AdS$_7$
solutions.

\bigskip
\noindent\underline{\bf $S^3$ bundle over $S^4$:}
\medskip

    We now turn our attention to the Euclidean theory.
In three dimensions, there exists a large class of squashed $S^3$ or
AdS$_3$ \cite{cps}.  We expect the same in seven dimensions. Without
loss of generality, we set $\Lambda=30$ so that it can give rise to
a unit round $S^7$.  We first consider the squashed $S^7$ that can
be viewed as an $S^3$ bundle over $S^4$.
The metric ansatz is given by%%%
%%%
\begin{equation} ds^2 = \alpha \sum_{i=1}^3(\sigma_i -
\cos^2(\ft12\theta)\,\tilde \sigma_i)^2 + \beta \Big(d\theta^2 +
\ft14\sin^2\theta \sum_{i=1}^3 \tilde
\sigma_i^2\Big)\,.\label{s3s4metric} \end{equation} %%%
%%%
where $\sigma_i$ and $\tilde \sigma_i$ are the $SU(2)$
left-invariant 1-forms, satisfying $d\sigma_i = \ft12 \epsilon^{ijk}
\sigma^j\wedge \sigma^k$ and $d\tilde \sigma_i = \ft12
\epsilon^{ijk} \tilde \sigma^j\wedge \tilde \sigma^k$. The metric is
Einstein provided that either $\alpha=\beta=\ft14$ or
$\alpha=\ft15\beta=\ft{9}{100}$. The first case corresponds to the
round $S^7$ and the second is a squashed $S^7$ that is also
Einstein. Now with the contribution from the topological terms, the
EOMs can be reduced to
%%%
\begin{equation} 2 \alpha^2 + 4\alpha\,\beta (7 \beta-2) - \beta^2=0\,,
\label{s3s4eom1}\end{equation} %%%
together with %%%
%%%
\begin{equation} \sqrt{\alpha} (\alpha-\beta)^3 (4 (10\alpha + \beta)
\tilde \mu
-(55\alpha + 7\beta)\nu) + 2\beta^6 (20\alpha\beta -4\alpha -
\beta)=0\,. \label{s3s4eom2} \end{equation}
%%%
It is clear from (\ref{s3s4eom1}) that there exists one and only one
positive $\alpha$ for any positive $\beta$.  The squashing parameter
$\gamma\equiv\alpha/\beta$ lies in the range $0 <\gamma < 2 +
\fft{3}{\sqrt2}$. Note that when $2\tilde\mu=3\tilde\nu$, the squashed
$S^7$ that is Eisntein remains Einstein.

\bigskip
\noindent\underline{\bf $S^1$ bundle over $\CP^3$:}
\medskip

There is another way of squashing an $S^7$, which can be viewed as
an $S^1$ bundle over $\CP^3$.  This example can be generalized to
Minkowskian signature to give rise to squashed AdS$_7$
\cite{squashads}. The metric ansatz is given by %%%
%%%
\begin{eqnarray} ds^2&=&\alpha\, (d\tau + \sin^2\theta
(d\psi + B))^2 + \beta\, ds_{\CP^3}^2\,,\cr ds_{\CP^3}^2 &=&
d\theta^2 + \sin^2\theta\, \cos^2\theta (d\psi + B)^2 + \sin^2\theta
\Big(d\tilde \theta^2 + \ft14 \sin^2\tilde\theta\,
\cos^2\tilde\theta\, \sigma_3^2\cr &&\qquad\qquad +
\ft14\sin^2\tilde\theta\, (\sigma_1^2 + \sigma_2^2)\Big)\,,\cr
B&=&\ft12 \sin^2\tilde\theta\, \sigma_3\,. \end{eqnarray} %%%%
%%%
It is of a round $S^7$ when $\alpha=\beta=1$.  In
general, the EOMs imply that %%%%
%%%
\begin{equation} \alpha = \beta (8-7\beta)\,,\qquad
8\tilde \mu + \tilde \nu + \fft{\beta^3}{10976 (\beta-1)^2
\sqrt{\alpha}}=0\,.
\end{equation} %%
%%%
The squashing parameter $\gamma\equiv \alpha/\beta$ lies in the
range $(0,8)$.

\bigskip
\noindent\underline{\bf Squashed $Q^{111}$ spaces:}
\medskip

The $Q^{111}$ space is an Einstein-Sasaki space of $U(1)$ bundle
over $S^2\times S^2\times S^2$.  We consider the following ansatz
\begin{equation}
ds^2=\alpha \Big(d\psi + \sum_{i=1}^3 \cos\theta_i\, d\phi_i\Big)^2
+ \beta \sum_{i=1}^3 (d\theta_i^2 + \sin\theta_i^2 d\phi_i^2)\,.
\end{equation}
%%%
It is of $Q^{111}$ provided that $\alpha=\ft12\beta=1/16$, and it
remains so for $\tilde\nu=0$. In general, we have
\begin{equation}
\alpha = 4\beta (1-7\beta)\,,\qquad 8(\alpha -\beta) (2\alpha
-\beta) \tilde \mu + \alpha (2\alpha - 3\beta) \tilde \nu +
\fft{\beta^5(\alpha -
8\beta + 60\beta^2)}{4\alpha^{3/2}}=0\,.
\end{equation}
Thus the squashing parameter $\gamma\equiv\alpha/\beta$ lies in the
range $(0,4)$.  We expect that many of the squashed homogeneous
spaces in seven dimensions are now solutions in this new gravity
theory, and we shall not enumerate them further.

\section{Conclusions}

This work is motivated by studying the classical solutions of
Einstein-Chern-Simons gravity with asymptotic AdS structure. In
seven dimensions, there are two topological Chern-Simons terms, and
we obtain the full set of equations of motion.  We find that
spherically-symmetric solutions are unmodified by the inclusion of
these topological terms.  We also obtain squashed AdS$_7$, and
squashed $S^7$ and $Q^{111}$ spaces in Euclidean signature,
where the squashing parameter is related to the
coupling constants of the topological terms.  It is intriguing to
see that these known squashed homogeneous spaces which appear to
have no connection can now be unified under our new gravity theory.

     As in three dimensions, our topological gravity should play
an important role in exploring the AdS$_7$/CFT$_6$ correspondence.
The CFT$_6$ that describes the world-volume theory of multiple
M5-branes is yet to be known, and our solutions provide many new
gravity dual backgrounds.  The quantization condition for one of the
coupling constant suggests an unusual property of the CFT$_6$ that
is absent in lower dimensions. Additional future directions include
a classification of all topological gravities in $(4k+3)$
dimensions, investigating the linearization of $D=7$ topological
gravity and obtaining the propagating degrees of freedom.

\section*{Acknowledgement}

We are grateful to Chris Pope for useful discussions.  Y.P. is
supported in part by the NSFC grant No.1053060/A050207 and the NSFC
group grant No.10821504.

%%%%%%%%%%%%%%%%%%%%%%%%%%%%%%%%%%%%%%%%%%%%%%%%%%%%%%%%%%%%%%%%%%%%%%%


\begin{thebibliography}{99}


%\cite{Deser:1981wh}
\bibitem{Deser:1981wh}
  S.~Deser, R.~Jackiw and S.~Templeton,
  ``Topologically massive gauge theories,''
  Annals Phys.\  {\bf 140}, 372 (1982)
  [Erratum-ibid.\  {\bf 185}, 406.1988\ APNYA,281,409 (1988\ APNYA,281,409-449.2000)].
  %%CITATION = APNYA,281,409;%%

%\cite{Wilczek:1983cy}
\bibitem{Wilczek:1983cy}
  F.~Wilczek and A.~Zee,
  ``Linking numbers, spin, and statistics of solitons,''
  Phys.\ Rev.\ Lett.\  {\bf 51}, 2250 (1983).
  %%CITATION = PRLTA,51,2250;%%

%\cite{Carroll:1989vb}
\bibitem{Carroll:1989vb}
  S.M.~Carroll, G.B.~Field and R.~Jackiw,
  ``Limits on a Lorentz and parity violating modification of electrodynamics,''
  Phys.\ Rev.\  D {\bf 41}, 1231 (1990).
  %%CITATION = PHRVA,D41,1231;%%

%\cite{Deser:1983tn}
\bibitem{Deser:1983tn}
  S.~Deser, R.~Jackiw and G.~'t Hooft,
  ``Three-Dimensional Einstein gravity: dynamics of flat space,''
  Annals Phys.\  {\bf 152}, 220 (1984).
  %%CITATION = APNYA,152,220;%%

%\cite{Witten:2007kt}
\bibitem{Witten:2007kt}
  E.~Witten,
  ``Three-dimensional gravity revisited,''
  arXiv:0706.3359 [hep-th].
  %%CITATION = ARXIV:0706.3359;%%

\bibitem{stromingeretal}
  W.~Li, W.~Song and A.~Strominger,
  ``Chiral gravity in three dimensions,''
  JHEP {\bf 0804}, 082 (2008)
  [arXiv:0801.4566 [hep-th]].
  %%CITATION = JHEPA,0804,082;%%
%\cite{Bergshoeff:2009hq}
  E.A.~Bergshoeff, O.~Hohm and P.K.~Townsend,
  ``Massive gravity in three dimensions,''
  Phys.\ Rev.\ Lett.\  {\bf 102}, 201301 (2009) arXiv:
  0901.1766 [hep-th].
  %%CITATION = PRLTA,102,201301;%%

%\cite{Gunaydin:1984nt}
\bibitem{Gunaydin:1984nt}
  M.~Gunaydin, G.~Sierra and P.K.~Townsend,
  ``Quantization of the gauge coupling constant in a five-dimensional
  Yang-Mills/Einstein supergravity theory,''
  Phys.\ Rev.\ Lett.\  {\bf 53}, 322 (1984).
  %%CITATION = PRLTA,53,322;%%

%\cite{Pernici:1984xx}
\bibitem{Pernici:1984xx}
  M.~Pernici, K.~Pilch and P.~van Nieuwenhuizen,
  ``Gauged maximally extended supergravity in seven-dimensions,''
  Phys.\ Lett.\  B {\bf 143}, 103 (1984).
  %%CITATION = PHLTA,B143,103;%%

%\cite{Deser:2003vh}
\bibitem{Deser:2003vh}
  S.~Deser and B.~Tekin,
``Energy in topologically massive gravity,''
  Class.\ Quant.\ Grav.\  {\bf 20}, L259 (2003), gr-qc/0307073.
  %%CITATION = CQGRD,20,L259;%%

%\cite{Tachikawa:2006sz}
\bibitem{Tachikawa:2006sz}
  Y.~Tachikawa,
  ``Black hole entropy in the presence of Chern-Simons terms,''
  Class.\ Quant.\ Grav.\  {\bf 24}, 737 (2007), hep-th/0611141.
  %%CITATION = CQGRD,24,737;%%

\bibitem{cps}
  D.D.K. Chow, C.N. Pope and E. Sezgin,
``Exact solutions of topologically massive gravity,''
  arXiv:0906.3559 [hep-th].
  %%CITATION = ARXIV:0906.3559;%%

\bibitem{squashads}
  P. Hoxha, R.R. Martinez-Acosta and C.N. Pope,
``Kaluza-Klein consistency, Killing vectors, and Kaehler spaces,''
  Class.\ Quant.\ Grav.\  {\bf 17}, 4207 (2000), hep-th/0005172.
  %%CITATION = CQGRD,17,4207;%%

\end{thebibliography}
\end{document}